# FPGA with Improved Routability and Robustness in 130nm CMOS with Open-Source CAD Targetability

Guanshun Yu, Tom Y. Cheng, Blayne Kettlewell, Harrison Liew, Mingoo Seok, Peter R. Kinget

Department of Electrical Engineering
Columbia University
New York, New York, U.S.A
{gy2226,tyc2106,rbk2135,hl2670,ms4415,pk171}@columbia.edu

*Abstract*—This paper outlines an FPGA VLSI design methodology that was used to realize a fully functioning FPGA chip in 130nm CMOS with improved routability and memory robustness. The architectural design space exploration and synthesis capability were enabled by the Verilog-to-Routing CAD tool. The capabilities of this tool were extended to enable bitstream generation and deployment. To validate the architecture and bitstream implementation, a Chisel (Constructing Hardware in the Embedded Scala Language) model of the FPGA was created to rapidly verify the microarchitectural details of the device prior to schematic design. A custom carrier board and configuration tool were used to verify correct operational characteristics of the FPGA over various resource utilizations and clock frequencies.

## I. Introduction

The contribution of this paper is offering a complete, open-source FPGA design methodology capable of taping out an FPGA with Verilog programmability. This paper also discusses unique hardware design features that improve routability and memory robustness.

An FPGA is a reconfigurable integrated circuit that allows many classes of algorithms to be efficiently mapped to physical hardware resources. Its balance between performance, flexibility, and power efficiency has contributed to its increased adoption in modern computing applications, including digital signal processing and machine learning [1][2].

The FPGA prototype in this paper is designed using a semi-custom VLSI design flow; it includes 12 custom macro-blocks and synthesized memory decoders to access 26KB of programming SRAM. Along with the synthesized memory decoders, the FPGA was designed from the transistor level up using IBM 130nm bulk CMOS technology, and fabricated on a 2.25mm$^2$ die (Fig. 1).

VTR [3] is used to synthesize a Verilog input file to a set of output files, including Berkeley Logic Interchange Format (BLIF), netlist, and placement and routing file. These files are then parsed by a Scala based bitstream generator in order to create a binary file to configure the FPGA. The bitstream contents are organized into a hierarchical memory structure that consists of a 19×19 macroblock matrix with 9 locally accessible 8-bit words within each block. Individual memory cells enable routing paths via transmission gates, or store LUT values. The configuration bus (CDATA[0:7]) and its control signals are both asynchronous and parallel, borrowing heavily from the physical interface of asynchronous DRAM. This configuration approach simplifies timing analysis and allows rapid reconfiguration of the FPGA from the host device.

## II. Software Flow

Our augmented VTR CAD flow converts a Verilog input file to a binary file for our FPGA by going through four main stages as shown in Fig. 2. VTR only generates files with information about what each LUT contains and which routing tracks are active and provides no information about how to configure and interface with a physical FPGA. For that purpose, our Scala based bitstream generator tool parses the VTR outputs with knowledge of the custom FPGA microarchitecture. With VTR, designers are able to quickly explore a particular FPGA architecture, and validate it on a system level.

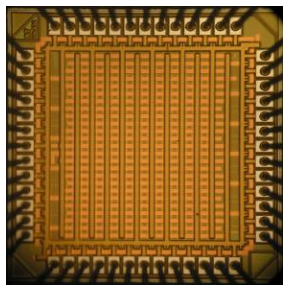

Fig. 1. Die Photo of The 130nm CMOS Custom FPGA Prototype

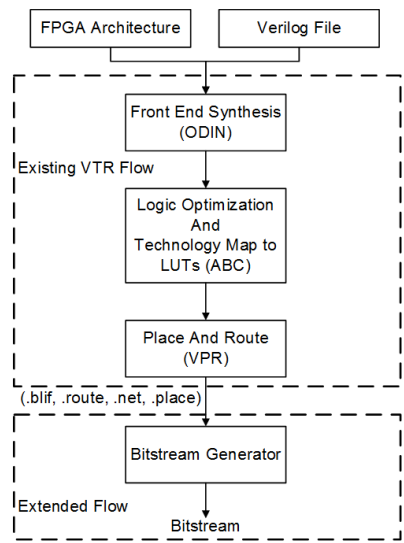

Fig. 2. Added Bitstream Generator On Top of Existing VTR Flow

To validate the chosen FPGA architecture, a behavioral model of the FPGA was constructed using Chisel [4], a new, hierarchical and object-oriented alternative to existing HDLs. Behavioral and structural Verilog models were verified using Modelsim. To better understand the true performance of the FPGA, the transistor-level design containing both custom and synthesized blocks was exported to SPICE netlists for simulation.

### III. Hardware Design

#### A. FPGA Chip Architecture

Our FPGA design is tile-based, so in theory, the design is expandable to an arbitrary size. Nevertheless, this particular prototype is limited to the pin count and die size specified from the MOSIS Educational Program. The design uses 52-pin QFP open cavity package and it has 16 GPIO ports and 8 Host Interface Ports (HIP). A top level block diagram is shown in Fig. 3.

Fig. 4. shows the architecture of one FPGA tile which includes the CLB, horizontal connection block (HCB), vertical connection block (VCB), and switch block (SB). To route a CLB's output to any other CLB's input, there are two types of routing blocks. Connection blocks connect a CLB's inputs and outputs to the routing grid. At the intersections of these routing tracks, switch blocks allow signals coming in from one direction to route to any of the other three directions.

LUT values and routing configuration in the FPGA core are stored in SRAM cells. An asynchronous parallel configuration bus loads the correct configuration values into all of the SRAM cells. Operating similarly to a DRAM bus, it has 8 bits of parallel data along with row and column decoders, so that a byte of SRAM at any given row and column can be selected and written. Moreover, the design allows for SRAM readback to detect any configuration errors and non-functional SRAM cells.

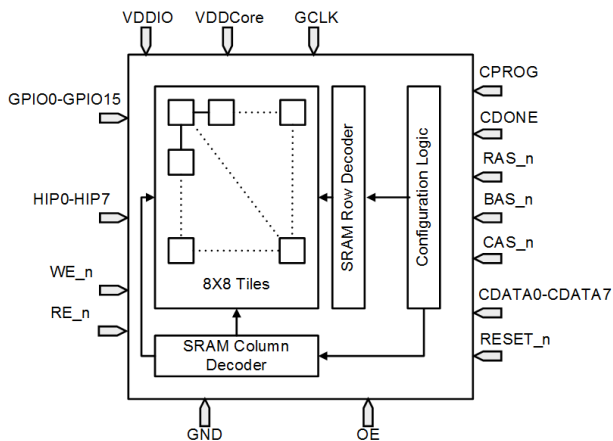

Fig. 3. Prototype FPGA Top-Level Block Diagram

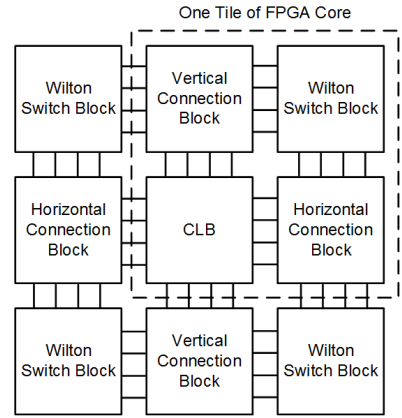

Fig. 4. Architecture of the FPGA Tile

#### B. Block Level Circuit Design

##### 1) Switch Block

The design of the routing architecture is critical, as routing resources consume most of the chip area and are responsible for most of the propagation delay in the FPGA [5]. Fig. 5 shows the three most common switch block topologies: Disjoint, Universal and Wilton. All of the switch blocks can connect each incoming track to three other outgoing tracks but the three different topologies have different input to output mapping.

The Wilton switch block topology, which offers high routability at the expense of some area [6], was chosen for this FPGA. Based on experimental results from [7], the Wilton switch block has a maximum of 28.57% routing channel reduction compared to other switch block topologies due to its overall switch flexibility. In terms of routability and minimum routing channel width, the Wilton switch block gives better results than the other two types of switch blocks.

The switches in this FPGA are implemented using transmission gates. Transmission gates avoid the need for a higher voltage power domain (and the associated package pins), unlike pass-gates that require gate boosting to compensate for the threshold drop across source and drain. Moreover, transmission gate based FPGAs were found to consume only 15% more area than pass transistor based design but to be 25% faster without gate boosting in [8].

##### 2) Configurable Logic Block

The FPGA's logic is comprised of 64 configurable logic blocks (CLBs) that each contain a 6-input LUT. The configurable logic block consists of a 6-stage MUX tree and memory that forms the LUT. A D flip-flop takes the output of the LUT and produces a synchronous output. The final 2-to-1 MUX can select either a synchronous or asynchronous output as shown in Fig. 6. In addition to the MUX tree and LUT, the design includes AND gates to mask unused input signals. Layout-extracted SPICE simulation at 25ºC and typical process corner shows that there is a 521ps critical path delay in the CLB.

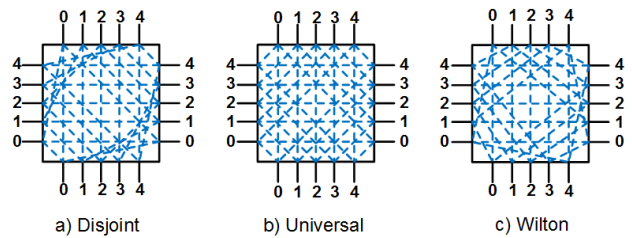

Fig. 5. Three Common Switch Blocks In Commercial FPGAs, Wilton Switch Block Is Used In Design

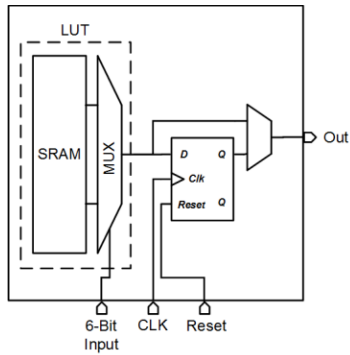

Fig. 6. Configurable Logic Block Architecture

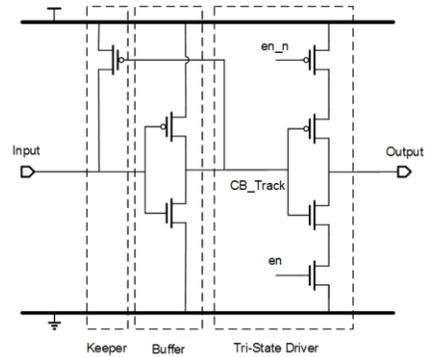

Fig. 7. One Track of The Connection Block

*3) Connection Block*

The connection block serves as the interface between CLBs and the routing grid. In addition, as shown in Fig. 7, each track inside the connection block contains a tri-state driver to prevent contention between signal lines and half-latches to remove floating nodes in unused routing tracks.

*4) I/O Block with 1.2V-3.3V Level Conversion*

The FPGA is fabricated using 130nm CMOS technology with a 1.2V core supply. To interface with external 3.3V LVCMOS signals, the FPGA's I/O block (Fig. 8) employs level conversion. Generally, contention is present in common level-up converters, degrading performance and increasing power consumption. The chosen level-up converter [9] avoids contention for up to 500MHz conversion speed and has 24mA of drive strength with slew rate control. A Schmitt trigger at the input rejects noise on the bus, and a bus keeper replaces configurable pull-up/down circuits. According to extracted SPICE simulations at 25ºC and typical process corner, level-up conversion has approximately 2.2ns of delay with a 10pF loading and level-down conversion [10] has 1ns of delay with 50fF capacitive loading.

*5) Clock Distribution Network*

To maintain clock signal integrity and minimize skew, a combination of the H-tree and grid-based clock distribution architectures is implemented in the FPGA, as shown in Fig. 9. In addition, ground shields are placed between the H-tree and surrounding clock grid. SPICE simulations at 25ºC and typical process corner show a rise and fall time of 300ps with this design. The propagation delay from the H-tree input to internal clock grid is 1.2ns. Simulation performance data also suggests that the core max clock frequency is 250MHz.

*C. SRAM Cell Design*

For an SRAM-based FPGA, having an extremely reliable SRAM is critical. This FPGA design uses 12 transistors to implement the SRAM cell, for robust readability and writability. Furthermore, the bitlines of the 12T SRAM design do not need to be precharged, so such a design does not need to be clocked. As a result, the absence of clock circuitry for the SRAM reduces power consumption after the initial configuration. Without the need for precharging bitlines, SRAM cells can also be placed more optimally in the design. The obvious penalty in area as compared to the commonly-used 6T SRAM design is partially offset in the FPGA design through aggressive use of minimum spacing and three metal layers. A compact 12T SRAM layout (Fig. 10) with dimensions of 47λ×64λ was achieved, which consumes 14% less area than the 12T SRAM layout in [11].

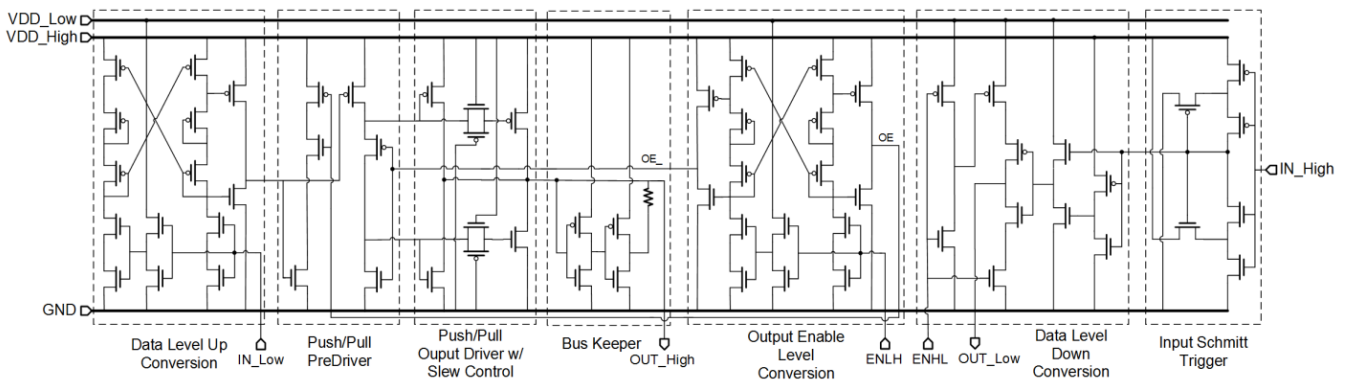

Fig. 8. Bidirectional Level Converter With Added Bus Keeper and Input Schmitt Trigger

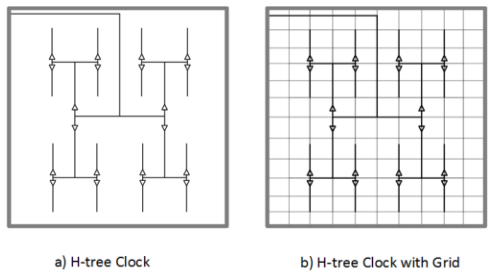

Fig. 9. H-tree with Grid Clock Architecture was Used

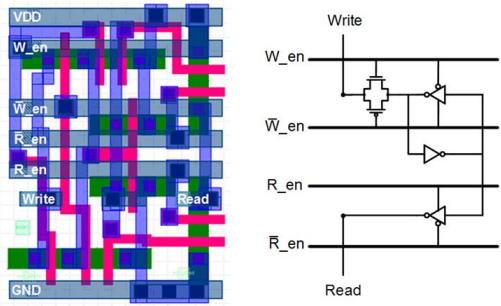

Fig. 10. Area-Optimized 12T SRAM Layout and Schematic

TABLE I. FPGA FREQUENCY AND POWER PERFORMANCE

| Application | Logic Utilization (64 CLBs Total) | Max Operating Frequency | Active Core Current At Max Frequency |
|---|---|---|---|
| *4 Bit Counter* | 4 CLBs | 100 MHz | 1380µA |
| *ASCII Encryption* | 24 CLBs | 80 MHz | 1150µA |
| *Gene Sequence Detector* | 43 CLBs | 80 MHz | 1040µA |
| *Stopwatch* | 60 CLBs | 40 MHz | 666µA |
| *Thermal Sensor* | 64 CLBs | 30 MHz | 491µA |

## IV. CHIP TESTING

This custom FPGA demonstrates successful design by correctly running arbitrary applications that are synthesized and routed from Verilog. According to test results shown in Table I, a higher logic utilization results in a lower maximum operating frequency. This is because more complex programs synthesize with longer critical paths, limiting the maximum operating frequency. Moreover, lower operating frequencies lead to lower active current consumption figures. All test applications met timing at a clock rate of at least 30MHz. A maximum clock rate of 100MHz was achieved using a synthesized 4-bit counter design (Fig. 11). FPGA core static leakage current of 123µA (at a nominal 1.2V core supply) was observed. The thermal sensor application demonstrates this architecture can utilize 100% of its logic resources.

## V. CONCLUSIONS

The FPGA design presented in this paper is a physical realization of a homogeneous FPGA with a Wilton switch block designed using the open-source VTR CAD flow. Moreover, the utilization of the Chisel language to rapidly model and test the FPGA design prior to schematic creation and synthesis demonstrated the value of this new HDL for implementing complex digital designs. The FPGA and its underlying sub-designs all function as intended as judged from the successful implementation of several demos. Lastly, this FPGA is logic limited, not routing limited as demonstrated by programs that approach full logic utilization.

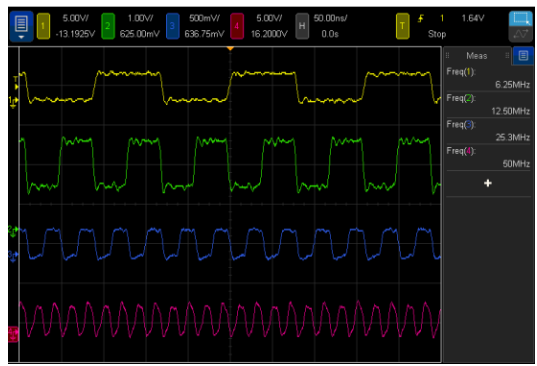

Fig. 11. 4-Bit Counter Waveform at 100MHz Core Frequency

### ACKNOWLEDGMENTS

The authors thank the MOSIS instructional program for fabrication and packaging support; the Electrical Engineering department for testing supplies; Vivek Mangal and Doyun Kim for technical assistance.